# Histo-fetch – On-the-fly processing of gigapixel whole slide images simplifies and speeds neural network training

*Brendon Lutnick[1], Leema Krishna Murali[2], Brandon Ginley[1], Avi Z. Rosenberg[3], and Pinaki Sarder[1,2*]*

[1]Department of Pathology & Anatomical Sciences, SUNY Buffalo, USA. [2]Department of Biomedical Engineering, SUNY Buffalo, USA. [3]Department of Pathology, Johns Hopkins University School of Medicine, Baltimore, USA.

*Address all correspondence to: Pinaki Sarder
Tel: (716)-829-2265, e-mail: pinakisa@buffalo.edu

## ABSTRACT
We created a custom pipeline (histo-fetch) to efficiently extract random patches and labels from pathology whole slide images (WSIs) for input to a neural network on-the-fly. We prefetch these patches as needed during network training, avoiding the need for WSI preparation such as chopping/tiling. We demonstrate the utility of this pipeline to perform artificial stain transfer and image generation using the popular networks CycleGAN and ProGAN, respectively.

## INTRODUCTION
A recent push for digitization in the field of pathology has created exciting opportunities for application of neural networks in diagnostic, prognostic, and theragnostic applications, beyond what could be accomplished with native histology image formats [1]. Whole slide images (WSIs) are high resolution scans of tissue sections, they are often giga-pixel sized and saved with multi-resolution compression formats [2]. This makes them too large to fit into the memory of hardware (i.e., GPUs) typically used to train convolutional neural networks (CNNs) [3]. To manage the data volume, WSIs are preprocessed by chopping them into patches which are fed to CNN architectures [3-9]. Traditionally chopping is performed before training, and pre-determined patch locations are saved to the disk as images. While this approach works, it is far from ideal and has limited scalability. Large WSI datasets are storage intensive, and chopping prior to training duplicates this data locally on the disk. To overcome this constraint, we developed histo-fetch, an innovative input pipeline for training CNNs on WSIs with the popular Tensorflow (tf) library [10]. Histo-fetch samples stochastic patch locations from WSI datasets actively during the network training, executing preprocessing and common data augmentation operations of this data on the CPU while the GPU simultaneously executes training operations (Fig. 1).

## RESULTS AND DISCUSSION
To demonstrate the usefulness of histo-fetch in training downstream deep learning training, we modified to two popular generative adversarial network (GAN) architectures, enabling direct implementation on WSIs. The first of these was the cycle-consistent adversarial network (CycleGAN) network [11], a popular architecture for translating between two image datasets. We perform artificial histological stain transfer using our modified CycleGAN, learning to map between stains. This network is extremely simple to train, requiring two WSI datasets (in this case with different stains) to be placed in separate folders. This network used 256 x 256 pixel patches downsampled to 1/4 resolution. An in-depth details of training are available in our prior work [12].

We trained two versions of this network using three different histological stains. Fifty-nine Silver and 313 hematoxylin and eosin (H&E) stained WSIs were each used as inputs for the two separate networks which learned to transfer their input to an *in silico* Periodic acid–Schiff (PAS) stain. Fifty-nine PAS WSIs were used by both the networks for the training. Examples from these stain transfer networks are visualized in Fig. 2A. The generated PAS images preserve the tissue structure of the input. Overall the results look promising. *In silico* silver and H&E images show correctly mapped basement membranes, and tubular brush borders. However, looking closely, the network maps H&E red blood cells into PAS nuclei.

Others have applied CycleGAN for histological stain translation, but generally use inefficient WSI chopping before network training [13, 14]. Thomas de Bel *et al.* [15] in their work mention a random sampling approach seemingly similar to our method, however, they do not provide any details of their algorithm, and their code is not publicly available. We propose that our methodology can be readily used for stain translation across color spaces that are of utility to pathologists in routine practice, without incurring an additional computational expense.

In the second application of our preprocessing method, we adapted NVIDIA's progressive growing of GANs (ProGAN) architecture [16]. To generate realistic looking *in silico* images, this network progressively grows the resolution of generated images as the training progresses. We trained the modified network up to an image size of 512 x 512 pixels. Macroscopically the *in silico* images look very realistic, exhibiting natural histological patterns from different stains. Looking closely at the image microanatomy, unexpected morphologies such as fused tubules, thick mesangium, and poorly generated glomeruli and tubule brush borders can be observed. We believe that further training would resolve these issues but this is outside the scope of this paper. Examples of generated image patches are displayed in Fig. 2B, with more results available via the link discussed under *Code & Data Availability* section.

Due to the nature of the progressive training, ProGAN requires a multi-resolution dataset. In general, this dataset is prepared by saving training images at all the intermediate resolutions. Instead, we use histo-fetch to extract these patches at different downsampled resolutions as needed during training on-the-fly. The dataset used to train this network contains 1331 WSIs totaling 765 GB disk-space of data. Without histo-fetch, training at this large-scale would not be possible given the computational resources available typical to standard research labs. This would require extracting all WSI patches and redundantly saving them at resolutions of 4, 8, 16, 32, 64, 128, 256, and 512 square pixels, requiring a massive use of disk storage. We hypothesize the diverse tissue morphology present in the large WSI training-set directly contributed to the quality of the synthetic images this network generates.

Training with histo-fetch is as simple as passing the folder location of the WSI dataset to the network. To quantify the efficiency of histo-fetch, we chopped a dataset of 151 human biopsy WSIs (18.9 GB) in the traditional method, saving the images to disk. Only regions containing tissue were saved as jpeg images (1200 x 1200 pixels) without overlap. This process took 5 hrs. 2 min. 57 sec. to complete, requiring 15.3 GB of additional disk space. Using more common methods, which save overlapping patches would be even more inefficient and more redundant [8]. In comparison, histo-fetch was 98.6% faster to start training and used 7535x less disk space (Fig. 1C). From an organizational standpoint, working natively with WSIs greatly streamlines the CNN training workflow using histological data. Files can be managed at the slide level simplifying trouble-shooting and data management. After training is complete histo-fetch is easily modified to extract predetermined patch locations for prediction on holdout WSIs.

One benefit of histo-fetch is the random selection of training patches. It is well known that random shuffling of data batches is beneficial during neural network training [17]. We argue that histo-fetch provides a more stochastic delivery of training data than pre-chopping WSIs. Without pre-determined data patch locations the network sees greater data variation during training. As a result, the notion of the training epoch (one training loop through all the data) is no longer valid, the number of training-steps should be specified instead.

Processing patches on the CPU at runtime increases the computational overhead, however training is GPU rate-limited. On our system, the CPU pre-fetches image patches faster than the GPU requests them. Training the two networks presented in Fig. 2 with a 10 core Intel Xeon (R) Silver 4114 CPU and Nvidia Quadro RTX 5000 GPU, we found no significant difference in GPU utilization or the execution speed of training steps when compared to using pre-extracted image patches. We quantified this using the ProGAN network, the average training-step was $0.813 \pm 0.005$ sec when chopping patches and encoding them into the default TFRecord format [10]. Updating ProGAN to use histo-fetch the training-step time was $0.819 \pm 0.006$ sec (Fig. 1C). Despite the 0.006 sec per step penalty, our ProGAN trained for 1.6M steps using histo-fetch (without WSI chopping and prep of TFRecord files) would still be much faster.

To conclude, we have developed a method for on-the-fly extraction and processing of patches from WSIs during neural network training. Using this method does not affect the speed of the training, but greatly reduces the time and disk-space requirements of dataset preparation prior to training. As an added benefit, our method provides greater randomness of data during training than the traditional method which is beneficial for network training [17]. Finally, we believe that once set up, this method is easier to use for novices and experienced data-scientists alike. WSI datasets are intuitively managed at the slide level, and changes do not require re-preparing the dataset. We believe that any CNN trained on gigapixel scale WSIs would benefit from the use of this method.

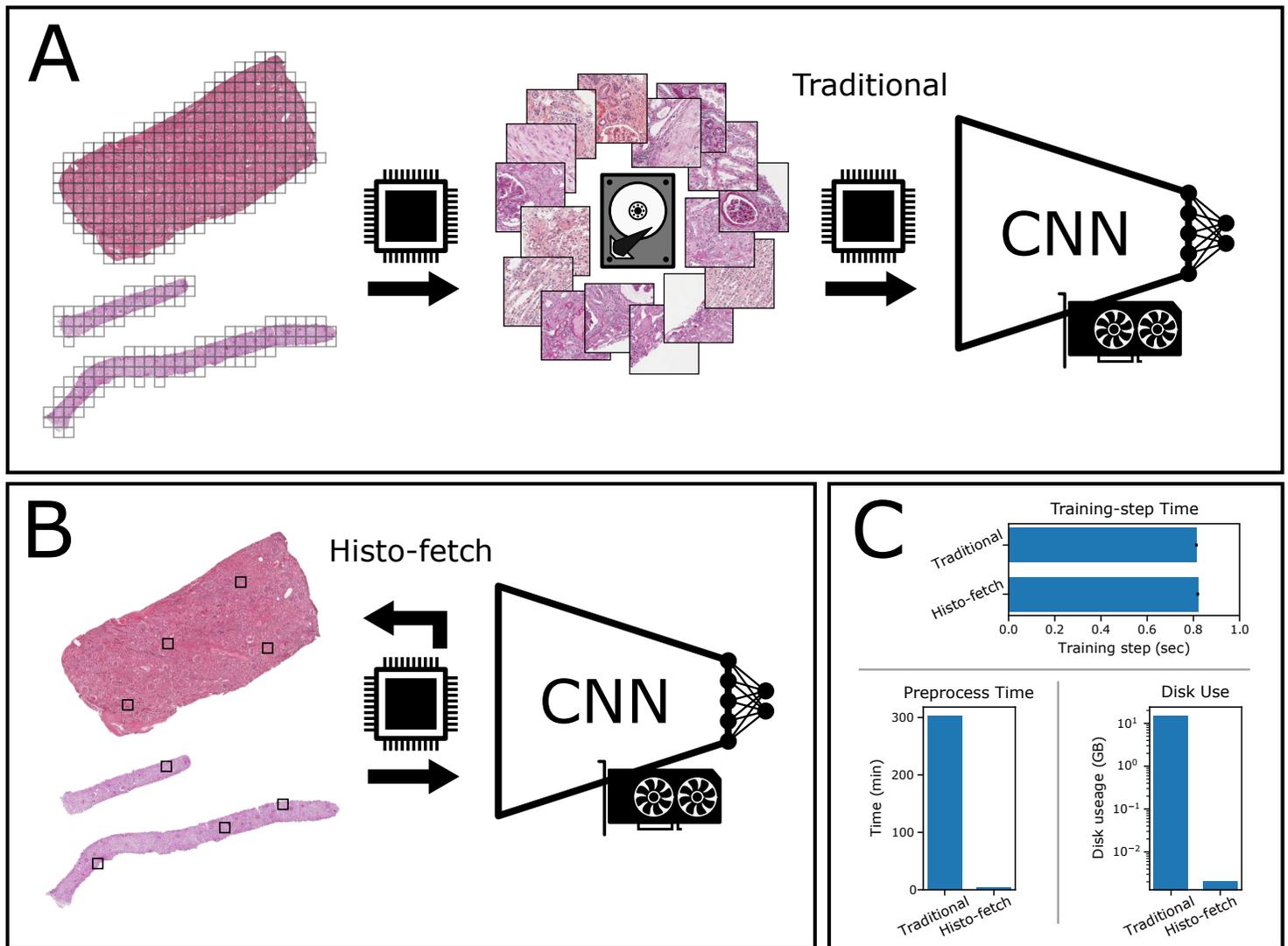

*Figure 1* | A comparative look at methods for training CNNs using WSI data.

**[A]** the traditional method. This uses the CPU to chop WSIs into patches which are saved to disk prior to CNN training. These patches are then read by the CPU at training time and fed to the GPU for training. **[B]** an overview of histo-fetch - our optimized method. We randomly select indices of patches that contain tissue at runtime. These patches are processed on the CPU and supplied to the GPU at training time. By prefetching the data, the CPU is always ready to supply the GPU with data as needed. **[C]** A comparison of efficiency between the two approaches. This highlights the preprocessing time before starting training and additional disk space required by both methods using a dataset of 151 human biopsy WSIs. Using ProGAN the average training step time does not significantly change using Histo-fetch.

## METHODS

Histo-fetch first does a pre-segmentation of the WSIs to identify the tissue regions using morphological image processing. A combination of blurring, thresholding via the Otsu method [18], and binary erosion identifies tissue and background regions. To maintain efficiency, we utilize the multiresolution decoding ability of WSI formats, using the thumbnail resolution for this pre-segmentation. The low-resolution tissue mask images are saved to the disk as an extremely space efficient 2-bit portable network graphics (PNG) file. We use the Tensorflow data.Dataset class to setup the input pipeline [10]. Specifically, a custom python function generates random coordinates within the previously identified tissue region, which are then loaded using the OpenSlide-python library [2]. This code takes arguments for the size and down-sampling of patches and has the ability to augment the patches via color-shifting, piecewise affine transformation, and rotation/flipping at runtime. For efficiency, we capitalize on the pyramidal multi-resolution encoding of WSIs when training with downsampled image patches, using the optimal slide resolution level for extraction using OpenSlide. This python function wrapped in a Tensorflow py_function operation which allows it to be executed as part of the network graph. Batching the image patches is handled by tf.data.Dataset, whose prefetch operation allows the CPU to preemptively prepare data batches during training.

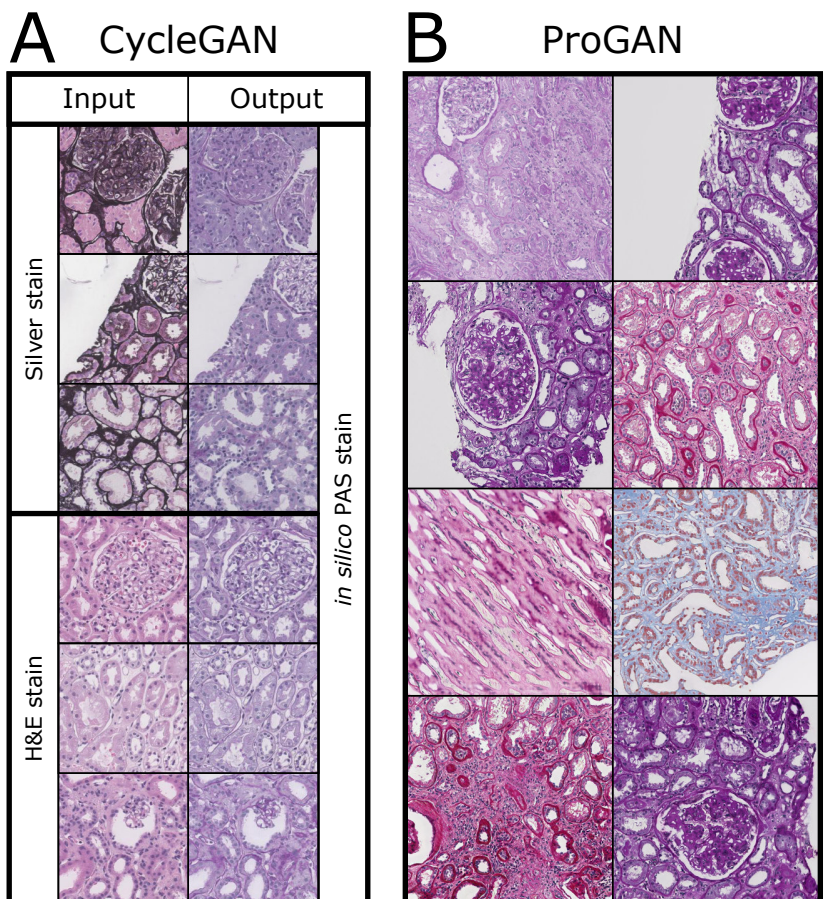

**Figure 2** | Example images from GAN architectures trained using histo-fetch.

**[A]** shows results from two CycleGAN networks, which take H&E or silver stained input patches and transform them to in silico PAS stains. **[B]** shows synthetic tissue patches generated using ProGAN trained on 1331 human biopsy (765 GB) WSIs with various histological stains.

## CODE & DATA AVAILABILITY

Histo-fetch and modified GANs codes with trained models are available at:
https://github.com/SarderLab/tf-WSI-dataset-utils
https://github.com/SarderLab/WSI-cycleGAN
https://github.com/SarderLab/WSI-ProGAN

ProGAN generated *in silico* images are available at: https://buffalo.box.com/s/ra5gp06kwcadpd9cefnqq0p103utip9x

## ACKNOWLEDGEMENT

The project was supported by the NIDDK grant R01 DK114485 & DK114485 02S1 and NIDDK Kidney Precision Medicine Project grant U2C DK114886.

## AUTHOR CONTRIBUTIONS

B.L. was responsible for conceptualizing and coding Histo-fetch, WSI-CycleGAN, and WSI-ProGAN as well as writing the paper. L.M. used the WSI-CycleGAN code to generate the results in Fig. 2A. B.G. assisted in manuscript preparation. P.S. was responsible for the overall coordination of the project, mentoring, and formalizing the image analysis concepts, and oversaw manuscript preparation.